\documentstyle[aps,prl,epsfig,floats,twocolumn]{revtex}
\begin{document}
\title{Shot noise in frustrated single-electron arrays}
\author{Daniel M. Kaplan, Victor A. Sverdlov, and Konstantin K. Likharev}
\address{
Department of Physics and Astronomy, State University of New York,
Stony Brook, New York 11794-3800 }
\date{\today }
\maketitle

\begin{abstract}
We have carried out numerical simulations of shot noise in 2D arrays of single-electron islands with random background charges. The results show that in contrast with the 1D arrays, at low currents the current noise is strongly colored, and its spectral density levels off at very low frequencies. The Fano factor may be much larger than unity, due to the remnants of single-electron/hole avalanches. However, even very small thermal fluctuations reduce the Fano factor below 1 for almost any bias. 
\end{abstract}


\bigskip

	Arrays of small conducting islands separated by tunnel junctions are among the simplest  systems exhibiting the Coulomb blockade and the related effects of correlated single-electron tunneling - for reviews see, e.g., Refs. \cite{ref:AveLik,DelRev,ref:Mooij}. Besides the successful use of multi-junction arrays for absolute thermometry (see, e.g., Ref. \cite{ref:4.31}), two their important applications in future digital nanoelectronics have been suggested:

	(i) The arrays may replace the double junctions in single-electron transistors, in order to raise the single-electron addition energy and hence increase the operation temperature range \cite{ref:4.21}. A quantitative analysis of this opportunity [6] has shown that for this purpose, arrays with zero background charges of the islands may be especially effective. 

	(ii) The arrays may serve as circuit elements with "sub-electron" (quasi-continuous) charge transfer \cite{ref:Matsuoka}, capable of leaking to ground the random background charges of single-electron islands induced by charged impurities - see, e.g., Ref. \cite{ref:Review2}. Earlier we showed \cite{ref:ZCB} that both conditions of sub-electron transfer, that had been formulated in the introduction to Ref. \cite{ref:Matsuoka}, are satisfied if the background charges of the array islands are arranged in a special way.

	Both applications are, however, strongly affected by the background charge randomness. In our recent work \cite{ref:4.32} we have shown that in the realistic case of "full charge frustration", when random charges are spread uniformly within the available range $[-e/2, +e/2]$ \cite{ref:4.33}, the former application becomes impossible, because the array loses any sensitivity to the gate voltage. Concerning the second application, the results of the analysis \cite{ref:4.32} were somewhat more encouraging: they showed that the background charge randomness suppresses the Coulomb blockade considerably, albeit not fully. Moreover, in large 2D arrays the statistical distribution of the blockage threshold voltage $V_t$ becomes relatively narrow, so that dynamic properties of these arrays are virtually reproducible sample-to-sample. In the light of this result, it was important to check whether the second condition of sub-electron transfer \cite{ref:Matsuoka}, the shot noise suppression, is also satisfied  in such arrays. 
\begin{figure}[tbp]
\centerline{\hbox{
\psfig{figure=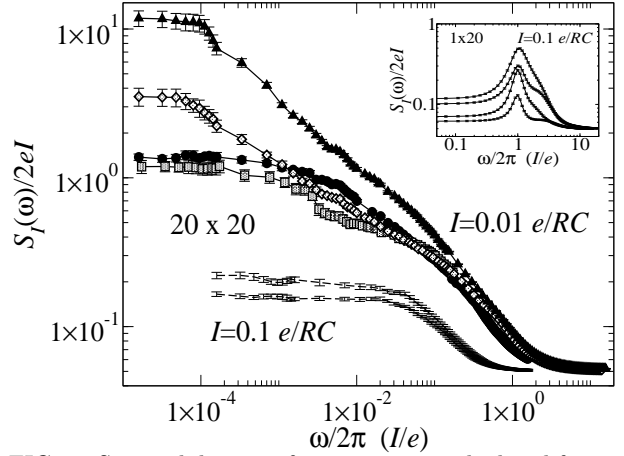,width=8cm}}}
%\vspace{0.2cm}
\caption{Spectral density of current noise calculated for several 2D arrays with random background charge distribution, as a function of observation frequency $\omega$, for two values of dc current $I$. For $I=0.1 e/RC$ only two extreme cases are shown. The inset shows the corresponding results for 1D arrays}
\label{fig:Ch4FvsOmg}
\end{figure}

\begin{figure}[tbp]
\centerline{\hbox{
\psfig{figure=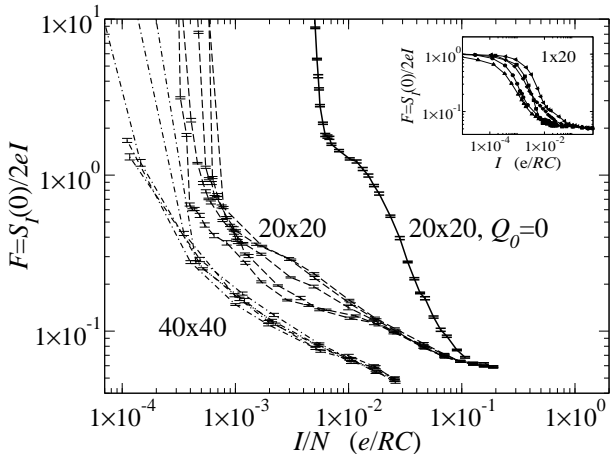,width=8cm}}}
%\vspace{0.2cm}
\caption{The Fano factor $F \equiv S_I(0)/2eI$ as a function of dc current per row for charge-frustrated arrays of two different sizes, and for the $20 \times 20$ array with zero background charge [20].
%\cite{ref:4.35}.
The inset shows the corresponding plots for frustrated 1D arrays.}
\label{fig:Ch4FvsI}
\end{figure}

	Unfortunately, despite the recent successful experimental implementation of ordered 2D arrays, whose transport properties are well described by the full-frustration model, by several groups (see, e.g., Refs. 12-16), we are not aware of any measurements of their noise properties. In this situation, the objective of this work was a numerical simulation of shot noise in 2D arrays with random background charge distribution, within the framework of the "orthodox" theory of single-electron tunneling \cite{ref:AveLik}. For comparison, results for 1D arrays are also shown.  (Some preliminary results for the 1D case were published earlier \cite{ref:Matsuoka}.) For the sake of simplicity we have studied the most interesting case of small island self-capacitance: $C_0 \ll C/N^2$, where $C$ is the capacitance between neighboring islands, and $N$ is the array size. The 2D arrays were of an almost square shape, with $N - 1$ islands along the current direction and $N$ islands across. Our algorithms were based on the well-known Monte Carlo code MOSES 1.2 that is available on the Web at http://hana.physics.sunysb.edu/set/software/index.html. Despite the code optimization, getting reasonable accuracy of noise calculations has required substantial supercomputer resources. For example, the calculation of just one group of curves shown in Fig. \ref{fig:Ch4FvsI} took approximately 1,500 node-hours of modern CPUs.

	Figure \ref{fig:Ch4FvsOmg} shows typical frequency dependences of the spectral density $S_I(\omega )$ of current noise in 2D arrays, for two values of dc current $I$ (determined by dc voltage $V \gtrsim V_t$ applied to the array) at zero temperature. At large currents and/or high frequencies the results are virtually independent on the system dimensionality; in fact, general arguments \cite{ref:Matsuoka,KorHop} show that in all cases $S_I(\omega )$ tends to $2eI/N$. However, at finite currents the results for 2D and 1D cases are quite different. In contrast to the 1D case, 2D arrays typically do not exhibit the spectral density increase at $\omega /2\pi = I/e$, corresponding to the so-called SET oscillations \cite{ref:AveLik}. More importantly, as dc current is reduced and/or the array size is increased, the frequency at which  $S_I(\omega )$ levels off decreases rather dramatically \cite{ref:4.34}, while the saturation level $S_I(0)$ increases well beyond the Schottky formula value $2eI$.  (In 1D arrays, $S_I(0)$ is always below $2eI$, so that the Fano parameter $F \equiv S_I(0)/2eI$ is always below unity \cite{ref:Matsuoka}.) Figure \ref{fig:Ch4FvsI} shows the Fano factor as a function of dc current for several arrays with random background charge, and for one $20 \times 20$ array with zero background charge. At $I \rightarrow 0$, the Fano factor saturates for any particular system, but at a very high level, with exponentially broad statistical distribution.

\begin{figure}[tbp]
\centerline{\hbox{
\psfig{figure=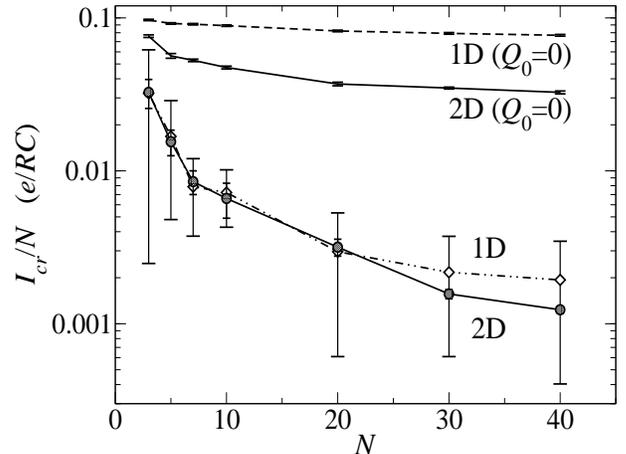,width=8cm}}}
%\vspace{0.2cm}
\caption{The average "crossover" current $I_{cr}$ (the dc current corresponding to $F = 1/N^{1/2}$) as a function the array length for 1D and 2D arrays, both with and without charge frustration. Error bars correspond to the r.m.s. width of the statistical distribution of $I_{cr}$.}
\label{fig:Ch4IcvsN}
\end{figure}

	Note that the dramatic increase of noise in 2D arrays is limited to low currents. This is why it does not affect too much such measure of the shot noise suppression as the "crossover" value $I_{cr}$ of dc current (Fig. \ref{fig:Ch4IcvsN}), at which the Fano factor drops to $1/N^{1/2}$, i.e. to the geometrical mean between the unity and its asymptotic, high-current value \cite{ref:Matsuoka}.

	The difference between results for 2D and 1D arrays may be attributed to the different topology of charge carrier motion. In the 1D case, current is carried along the array by single stream of single electrons (holes). Without disorder, the array provides a single bottleneck for this motion. At $I \rightarrow 0$, overcoming this bottleneck becomes a dominating source of current fluctuations, so that the current fluctuations are essentially the same as in a single tunnel junction, with a broadband shot noise and $F \rightarrow 1$ \cite{ref:Matsuoka}. Disorder in a long array induces multiple quasi-bottleneck states with very close thresholds $V_t^{(i)}$. The actual threshold voltage $V_t$ is determined by the highest of these thresholds, so that at $V \rightarrow V_t$ ($I \rightarrow 0$) the Fano factor still tends to unity.

	On the contrary, broad 2D arrays may feature multiple streams of electrons and holes moving in opposite directions, that do not necessarily annihilate even in the absence of disorder, giving rise to long "avalanches" producing in particular very large noise ($F \gg 1$) \cite{Aval}. The disorder suppresses this effect (Fig. \ref{fig:Ch4FvsI}), but only partly.   

\begin{figure}[tbp]
\centerline{\hbox{
\psfig{figure=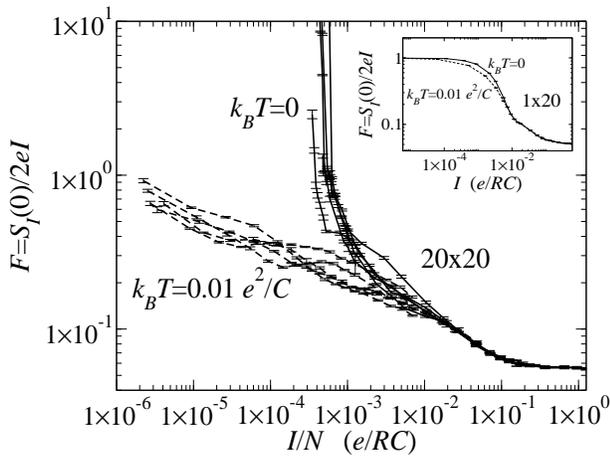,width=8cm}}}
%\vspace{0.2cm}
\caption{Fano factor as a function of dc current for several $20 \times 20$ arrays, at zero and finite temperatures. The inset shows the corresponding dependences for a typical charge distribution of a 1D array.}
\label{fig:Ch4FvsItemp}
\end{figure}

	It is interesting (and important for applications) that the Fano factor at low currents is substantially suppressed by even very small thermal fluctuations. For example, Fig. \ref{fig:Ch4FvsItemp} shows that temperature as low as $k_BT = 10^{-2}e^2/C$ reduces the Fano factor at low currents by several orders of magnitude, bringing it below unity, besides a very small region where $eV < 2k_BT$. (In that region, the current fluctuations are essentially thermal, rather than due to the shot noise.)  This behavior is very much different from that of 1D arrays (inset) where small increase of temperature have negligible effect on shot noise. 

Actually, Fig. \ref{fig:Ch4FvsItemp} may be somewhat misleading, because at vanishing temperature the limit $I \rightarrow 0$ corresponds to the voltage approach to a finite value $V_t$, while at $T > 0$ it is achieved at $V \rightarrow 0$, so we are speaking about very different dc bias regions.  In fact the decrease of $F$ is not  a result of the decrease of current fluctuations as such: actually at fixed applied voltage the spectral density $S_I(0$) always grows with temperature, but dc current grows even faster, resulting in the reduction of $F$. Nevertheless, Fig. \ref{fig:Ch4FvsItemp} relays the important information, since it is the Fano factor, not $S_I(0)$ per se, that is important for applications \cite{ref:Matsuoka}. 

	In fact, the reduction of $F$ by thermal fluctuations has been already noted in our work on avalanches in uniform arrays (see Fig. 5 of Ref. \cite{Aval}) and interpreted as avalanche suppression due to larger side diffusion of electrons and holes, increasing chances for their recombination. Apparently, disorder enhances this process and suppresses avalanches even further.  

	To summarize, we have found that large 2D single-electron arrays biased close to the Coulomb blockade threshold at $T \rightarrow 0$ (and hence carrying very low current) exhibit strongly colored noise, following approximately the $1/f^{\alpha}$ dependence until very low frequencies where it levels off yielding very high values of the Fano factor. This effect, that may be interpreted as a result of electron/hole avalanches in 2D arrays \cite{Aval}, would be very unfavorable for the application of such arrays as circuit components with quasi-continuous ("sub-electron") charge transfer. However, at small but finite temperatures the Fano factor is suppressed below unity for virtually all bias voltages. 
	
	Useful discussions with A. Korotkov are greatly appreciated. The work was supported in part by the Engineering Physics Program of the Office of Basic Energy Sciences at the Department of Energy and by the Semiconductor Research Corporation. We also acknowledge the use of following supercomputer resources: SBU's cluster {\it Njal} (purchase and installation funded by DoD's DURINT program), Oak Ridge National Laboratory's IBM SP computer {\it Eagle} (funded by the Department of Energy's Office of Science and Energy Efficiency program), and also two IBM SP systems: {\it Tempest} at Maui High Performance Computing Center and {\it Habu} at NAVO Shared Resource Center's (computer time granted by DOD's High Performance Computing Modernization program).


\begin{references}
\bibitem{ref:4.33}
The background charges may be defined modulo e because of the unavoidable thermal or dynamic annealing of the system \cite{ref:AveLik,DelRev,ref:Mooij,ref:Review2,ref:ZCB}.
\bibitem{ref:4.34}
One may say that at $I \rightarrow 0$ the infinite system cannot be characterized with the Fano factor at all, and the noise has a $1/f^{\alpha}$- type spectrum (in our results, $\alpha$ is crudely between 1/3 and 1/2). This effect is very close to that discussed for systems with variable-range hopping - see, e.g., Ref. \cite{ref:4.35} and citations therein.
\end{references}
\end{document}